\documentclass[12pt]{article}
\usepackage{graphicx}
\usepackage{color}
\usepackage{comment}
\usepackage{amsmath}
\usepackage{hyperref}

\def\Title#1{\begin{center} {\Large #1 } \end{center}}
\def\Author#1{\begin{center}{ \sc #1} \end{center}}
\def\Address#1{\begin{center}{ \it #1} \end{center}}

\newcommand\pubblock{\rightline{\begin{tabular}{l} \\
           \end{tabular}}}
\newenvironment{Abstract}{\begin{quotation}  }{\end{quotation}}
\newenvironment{Presented}{\begin{quotation} \begin{center} 
             PRESENTED AT\end{center}\bigskip 
      \begin{center}\begin{large}}{\end{large}\end{center} \end{quotation}}
\def\Acknowledgements{\bigskip  \bigskip \begin{center} \begin{large}
             \bf ACKNOWLEDGEMENTS \end{large}\end{center}}

\def\beq{\begin{equation}}
\def\eeq#1{\label{#1}\end{equation}}
\def\eeqn{\end{equation}}

\def\beqa{\begin{eqnarray}}
\def\eeqa#1{\label{#1}\end{eqnarray}}
\def\eeqan{\end{eqnarray}}

\let\bar=\overbar

\def\Dslash{\not{\hbox{\kern-4pt $D$}}}
\def\dslash{\not{\hbox{\kern-2pt $\del$}}}

\def\msb{{\bar{\ssstyle M \kern -1pt S}}}

\begin{document}
\begin{titlepage}
\pubblock

\vfill
\Title{Search for neutrinoless double-beta decay with SNO+}
\vfill
\Author{Vincent Fischer on behalf of the SNO+ collaboration}
\Address{University of California at Davis, Department of Physics, Davis, CA 95616, U.S.A.}
\vfill
\begin{Abstract}
The SNO+ experiment, located in SNOLAB, 2 kilometers underground in the Creighton mine, near Sudbury, Canada, is a large scale neutrino detector whose main purpose is to search for neutrinoless double-beta decay and thus probe the Majorana nature of the neutrino. 
With 780 tons of liquid scintillator loaded with tellurium, SNO+ aims at exploring the Majorana neutrino mass parameter space down to the inverted mass hierarchy region. 
The versatility of the SNO+ detector also allows it to detect solar and reactor neutrinos, provide a measurement of the geoneutrino flux, detect galactic core-collapse supernovae and perform nucleon decay searches. 
The SNO+ experiment is currently taking data with a detector fully filled with ultrapure water. 
The detector will be completely filled with liquid scintillator in the coming months and subsequently loaded with tellurium. 
\end{Abstract}
\vfill
\begin{Presented}
Thirteenth Conference on the Intersections of Particle And Nuclear Physics (CIPANP2018)\\
Palm Springs, U.S.A., May 29 -- June 3, 2018
\end{Presented}
\vfill
\end{titlepage}
\def\thefootnote{\fnsymbol{footnote}}
\setcounter{footnote}{0}
%

\section{Introduction}

SNO+ is a large scale multi-purpose liquid scintillator detector whose primary goal is to assess the Majorana nature of the neutrino through the observation of the, yet undetected, neutrinoless double-beta decay\,\cite{Andringa:2015tza}.
Located in the SNOLAB laboratory in Canada, SNO+ is the successor of the successful SNO experiment\,\cite{Jelley:2009zz} and utilizes parts of its infrastructure.

As described in Sections~\ref{sec:detector} and~\ref{sec:phases}, SNO+ will consist of three consecutive phases, each using a different detection medium and targeting different physics goals.
The final phase of the experiment will be focused on the detection of the neutrinoless double-beta decay process of $^{130}$Te. To this end, about 4~tonnes of natural tellurium will be loaded onto the 780~tonnes of liquid scintillator, already present in the detector since the previous phase.

In addition to this physics goals, the versatility of SNO+ will allow precise measurements of reactor and solar neutrinos, as well as geoneutrinos, detection of core-collapse galactic supernovae neutrinos and nucleon decay searches.

\section{The SNO+ detector}
\label{sec:detector}

The SNO+ detector is located in the Creighton mine in Sudbury, Ontario, Canada, at a depth of more than 2~km. This provides the detector with a rock overburden of about 6000~meters water equivalent, corresponding to a cosmogenic muon flux in SNO+ as low as 70~muons per day.\\

\begin{figure}[htb]
\centering
\includegraphics[width=0.5\textwidth]{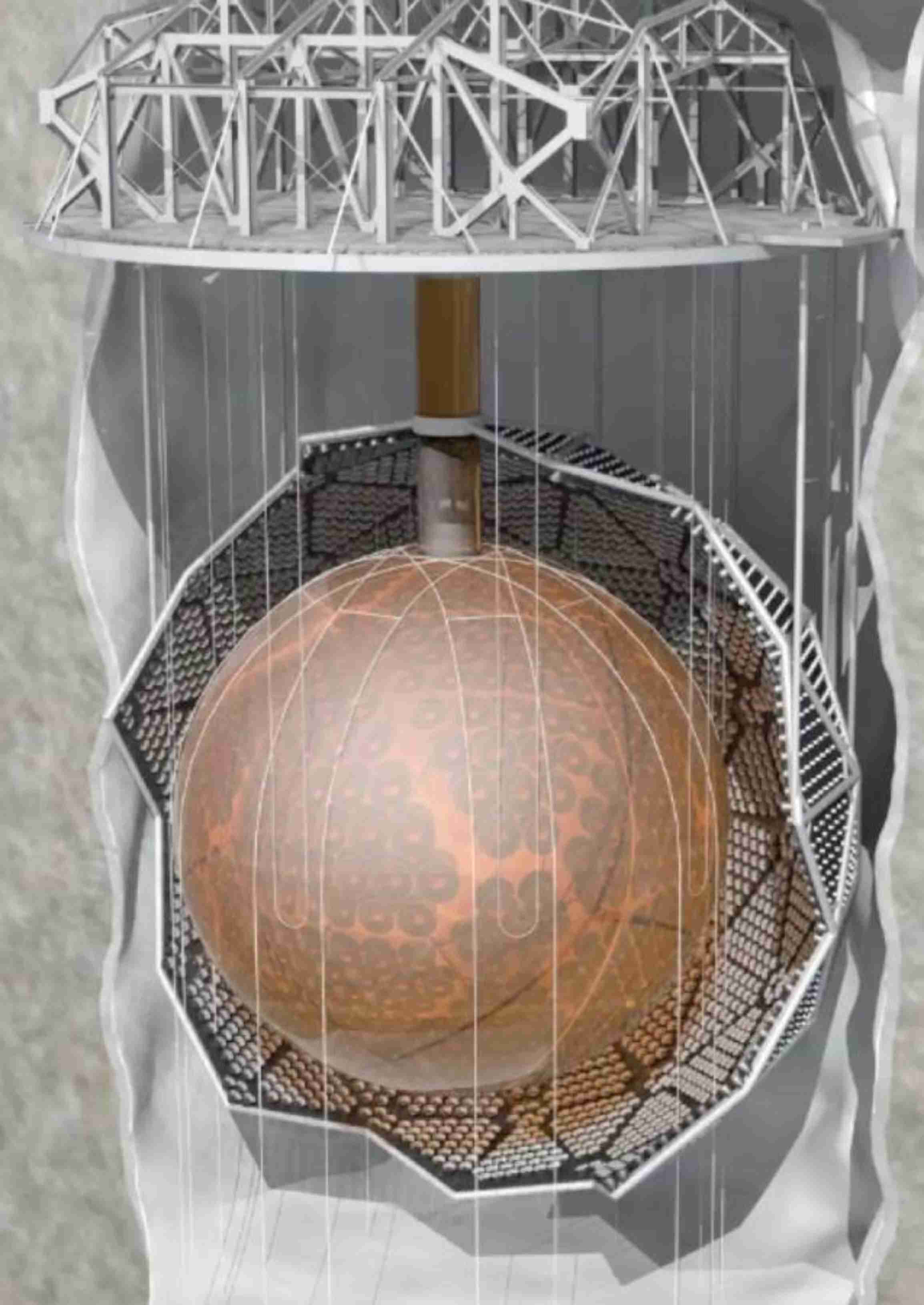}
\caption{Sketch of the SNO+ detector. The acrylic vessel (in brown) is held by hold-down and hold-up ropes and surrounded, in order, by the inner water shield, the PMT support structure, the outer water shield and the rock cavity. Above the detector lies the deck where the electronics and acquisition systems are located.}
\label{fig:detector_sketch}
\end{figure}

As displayed in Figure~\ref{fig:detector_sketch}, the detector is surrounded by an outer shield, a cavity in the rock filled with 5400~tonnes of ultrapure water, to protect it from the ambient radioactive background. 
A stainless steel support structure, with a diameter of 18 meters, holds the 9300~photomultiplier tubes (PMTs) of the detector, providing a photocoverage of about 53\%. 
While most PMTs are a legacy from SNO, a fair amount of them have been replaced or refurbished for SNO+. 
An inner shield, filled with 1700~tonnes of ultrapure water, prevents gamma rays from the radioactivity of the structure and the PMTs to reach the sensitive volume of the detector.

This inner volume, the acrylic vessel (AV), consists of a 5-cm thick acrylic vessel with a diameter of 12~meters. 
Previously used in SNO to hold heavy water, it had to be upgraded in order to hold liquid scintillator. 
Indeed, unlike heavy water, the density of liquid scintillator is inferior to the density of water (d$_\text{scintillator}$=0.86 g.cm$^{-3}$ at 12$^{\circ}$C in this case) thus causing a large buoyant force with respect to the surrounding water. A system of hold-down ropes has been installed to compensate this buoyancy and limit the stress on the acrylic vessel.\\

The acrylic vessel is now filled with ultrapure water for the first phase of the experiment (see Section~\ref{subsec:water_phase}).\\
For the second phase of the experiment (see Section~\ref{subsec:scintillator_phase}), the AV will be filled with a scintillator cocktail consisting of 780~tonnes of Linear Alkyl-Benzene (LAB), acting as a solvent, and 2~g/L of 2,5~diphenyloxazole (PPO), acting as a fluor. 
This LAB+PPO mixture has been chosen for its excellent compatibility with acrylic, high light yield (about 10,000 photons per MeV), high optical transparency, and good ability to discriminate between energy depositions from alpha particles and electrons.\\
Finally, for the third phase of the experiment (see Section~\ref{subsec:te_phase}), the scintillator cocktail will be loaded with 15~mg/L of bis-MSB, acting as a wavelength shifter, and 0.5\% $^{\text{natural}}$Te (by weight). 
This amount of natural tellurium corresponds to about 1330~kg of $^{130}$Te, the isotope of interest for double-beta decay searches.\\

In order to accommodate the higher light yield and lower energy threshold of the detector in its scintillator phase, the SNO electronics and data acquisition system have been upgraded for SNO+. 
Similarly, the calibration system has been upgraded to be compatible with liquid scintillator. 
This system now includes an array of optical fibers, coupled to LEDs and lasers, distributed throughout the entire detector. 
A light-diffusing sphere, as well as underwater cameras and a Cherenkov source, are also used to perform light calibrations and measure PMT and detection medium characteristics in-situ. 
To measure the efficiencies and systematics associated with energy and position reconstruction, various radioactive sources can be deployed along the vertical and horizontal axes of the detector.

\section{The Phases of SNO+}
\label{sec:phases}

\subsection{SNO+ Phase I - Water phase}
\label{subsec:water_phase}

The limited energy threshold and detection efficiency of the Cherenkov reaction in water does not allow SNO+ to tackle a wide range of physics.
Nonetheless, as SNO did some years ago, SNO+ can perform a measurement of the solar neutrino flux as well as nucleon decay searches. 
Using pure water instead of heavy water as a detection medium also allows the detector to be sensitive to electronic antineutrinos, emitted by nearby nuclear reactors, through the inverse beta decay reaction that generates a coincidence signal consisting of a positron and a neutron.\\

\begin{figure}[htb]
\centering
\includegraphics[width=0.8\textwidth]{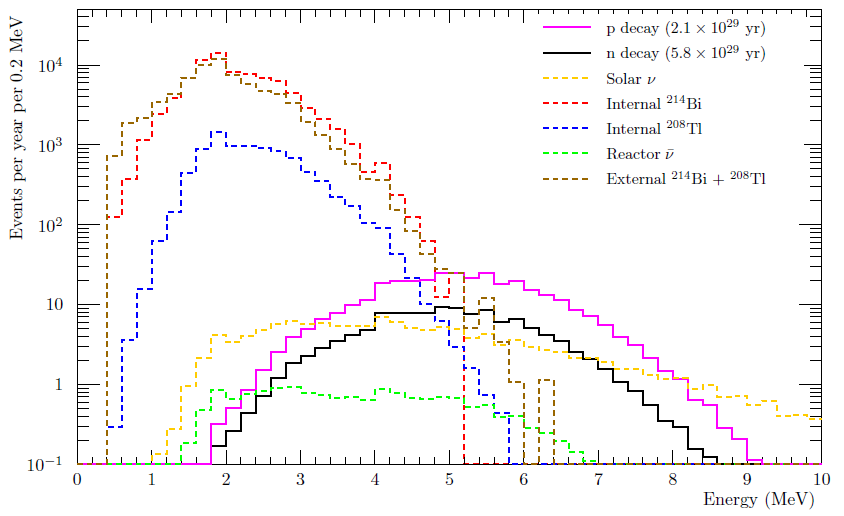}
\caption{Energy spectra of the expected signals (according to current limits) and backgrounds of the SNO+ water phase.}
\label{fig:nucleon_spec}
\end{figure}

The main physics goal of SNO+ in its water phase is the search for invisible nucleon decay in $^{16}$O. 
The sensitivity to those searches is further accentuated by the high overburden provided by the depth of SNOLAB, thus given SNO+ a considerable advantage over Super-Kamiokande.
A proton or neutron decay in $^{16}$O could lead to the creation of an excited $^{15}$N or $^{15}$O, respectively, that would subsequently de-excite through the emission of an energetic gamma ray (6 to 7~MeV)~\cite{Ejiri:1993rh}.
Such gamma rays could be easily distinguished above the ambient background in the SNO+ water as shown in Figure~\ref{fig:nucleon_spec}.\\
With 30 background events expected in the region of interest (ROI) after 6 months of data taking, SNO+ will be able to push the current limits on the invisible proton and neutron decay lifetimes up to $\tau_\text{n}$ $>$ 1.2 $\times$ 10$^{30}$ years and $\tau_\text{p}$ $>$ 1.4 $\times$ 10$^{30}$ years, thus improving the current limits set by KamLAND\cite{Araki:2005jt} ($\tau_\text{n}$ $>$ 5.8 $\times$ 10$^{29}$ years) and SNO\cite{Ahmed:2003sy} ($\tau_\text{p}$ $>$ 2.2 $\times$ 10$^{29}$ years) 

\subsection{SNO+ Phase II - Scintillator phase}
\label{subsec:scintillator_phase}

The lower detection threshold and higher light yield brought by scintillation light will allow SNO+ to tackle a broader range of physics goals.\\

SNO+ aims at measuring low-energy solar neutrinos, especially the \textit{pep} and \textit{CNO} contributions to the solar neutrino flux. 
While the detailed study of \textit{pep} neutrinos would improve our understanding of neutrino coupling to matter through oscillations at the transition between matter-dominated and vaccuum-dominated regions, the study of \textit{CNO} neutrinos would provide a measurement of the solar metallicity and help resolve the current discrepancy between recent measurements and predictions from the Standard Solar Model~\cite{Haxton:2008yv}.\\

Local nuclear reactors located around Toronto, Canada, will provide SNO+ with a flux of electronic antineutrinos that will result in about 100~events detected per year. 
By providing an independent measurement of the $\Delta m^{2}_{12}$ squared mass difference, SNO+ could help understand the existing tension between solar and reactor measurements performed by Super-Kamiokande and KamLAND, respectively~\cite{Abe:2016nxk}.
In addition, SNO+ expects to detect tens of geoneutrinos, emitted by the Earth's crust and mantle, hence providing a third measurement of the geoneutrino flux, after Borexino\cite{Agostini:2015cba} and KamLAND\cite{Araki:2005qa}, at a different location on the globe.\\

The scintillator fill is expected to start imminently and will last for about 3~months.
During this operation, the liquid scintillator, manufactured by CEPSA BECANCOUR in Quebec, will be shipped to SNOLAB in several weekly batches before being transfered underground using railcars. 
A dedicated plant has been built next to the detector for the purpose of further purifying the scintillator through distillation, metal scavenging, water extraction and nitrogen gas stripping to reduce the levels of uranium and thorium contaminations below 10$^{-17}$ and 10$^{-18}$ g/g, respectively.

\subsection{SNO+ Phase III - Tellurium phase}
\label{subsec:te_phase}

The third and final phase of SNO+ will be the tellurium phase whose main goal is to search for the neutrinoless double-beta decay of $^{130}$Te.
Unlike in a conventional double-beta decay, where the energy of the decay (the Q value) is shared between the two emitted electrons and the two emitted electronic antineutrinos, a neutrinoless double-beta decay is characterized by the fact that only electrons are found in the final state.
This results in a peak at the Q value of the reaction, instead of a continuous spectrum.
The number of observed decays, if any, provides the half-life of the process, directly related to the effective Majorana neutrino mass through the phase-space factor and the nuclear matrix elements of $^{130}$Te.\\

The SNO+ collaboration has chosen tellurium as a double-beta candidate for several reasons, including: the high abundance of $^{130}$Te in natural tellurium (34.08\%) thus eliminating the need for isotopic enrichment, the relatively high Q-value of $^{130}$Te ( 2527.01~keV) above most backgrounds and the possibility to load tellurium in liquid scintillator at high concentration while keeping low light attenuation.

The initial loading of SNO+ will be 0.5\% of natural tellurium by weight, corresponding to about 1300~kg of $^{130}$Te.
To load tellurium in liquid scintillator, an organo-metallic complex, telluric acid, is used along with a solvent, 1,2-butanediol.
Telluric acid has been purchased and stored underground for more than 3 years in order to let isotopes created upon exposure to cosmic rays (e.g. $^{60}$Co, $^{110m}$Ag, $^{88}$Y) decay as they are a source of background in the energy window of interest for double-beta decay searches.
Two dedicated plants have been built underground to filter and purify telluric acid as well as mix it with butanediol before loading the final tellurium complex into the detector.
With the help of this purification process, the levels of uranium and thorium contaminations are expected to be kept below 1.3 $\times$ 10$^{-15}$ and 5 $\times$ 10$^{-16}$ g/g, respectively.\\

\begin{figure}[htb]
\includegraphics[width=0.48\textwidth]{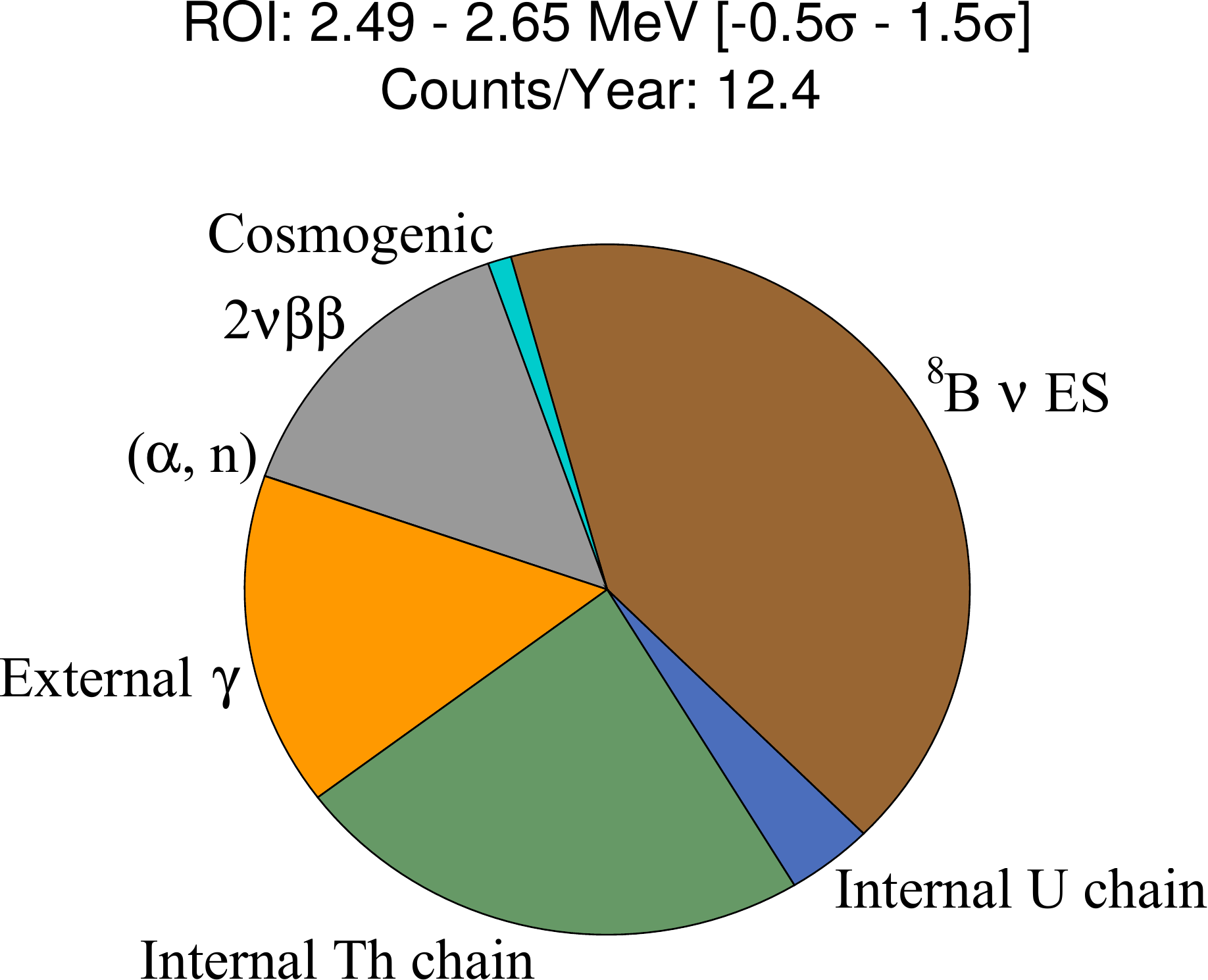}
\includegraphics[width=0.48\textwidth]{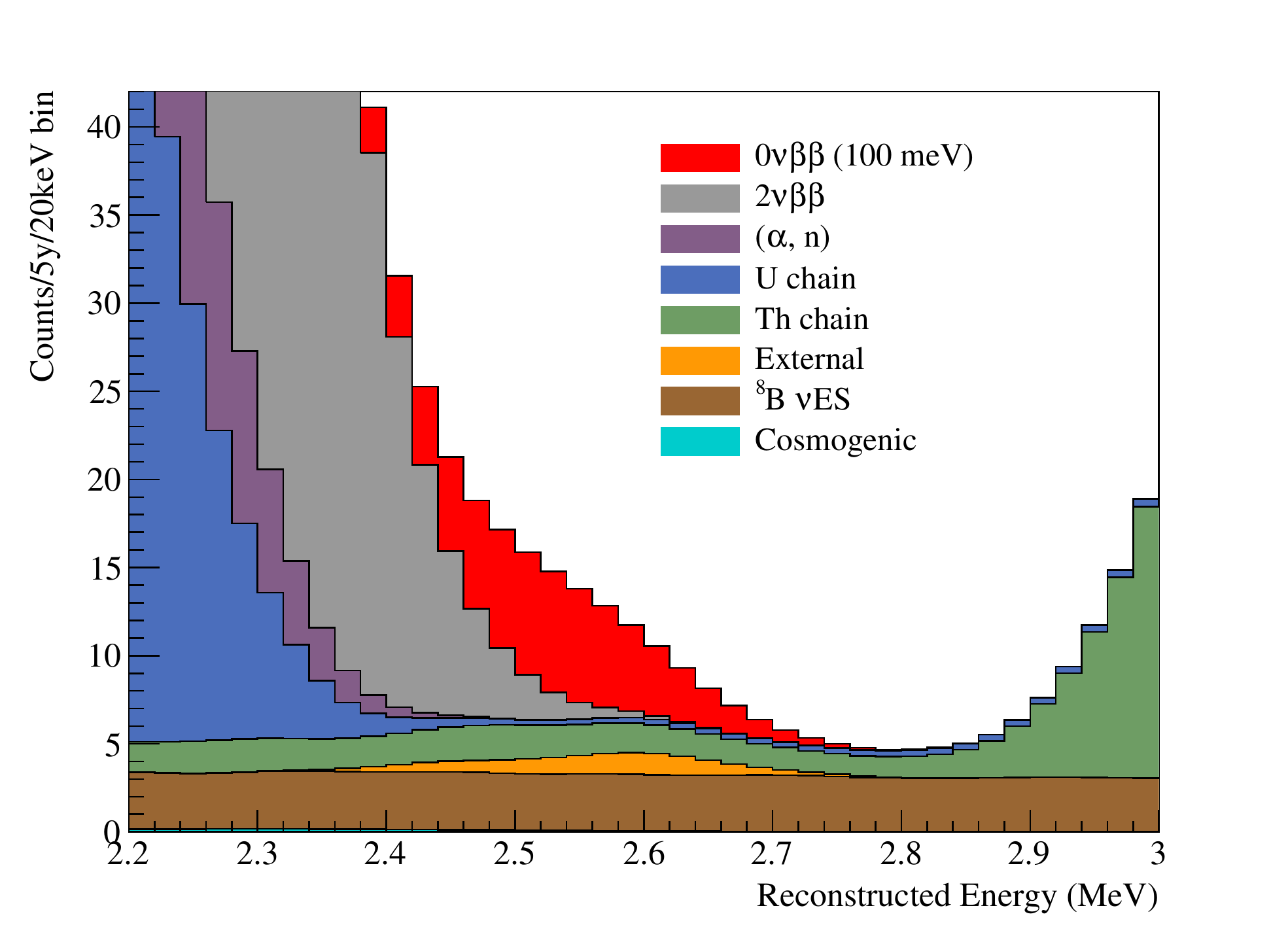}
\caption{Left: Background estimation for SNO+ tellurium phase. Right: Expected energy spectrum in the region of interest, showing all backgrounds and signal}
\label{fig:phase2_dbd}
\end{figure}

In the region of interest, centered on the $^{130}$Te Q-value and asymmetric [-0.5$\sigma$; +1.5$\sigma$] due to the large 2-neutrino double-beta background, expected backgrounds, shown in Figure~\ref{fig:phase2_dbd} (left), are:
\begin{enumerate}
\item[-] Elastic scattering of solar neutrinos from $^{8}$B
\item[-] Internal uranium and thorium contaminations
\item[-] External gamma rays (PMTs, rock, etc..)
\item[-] ($\alpha$,n) coincidence reactions
\item[-] 2-neutrino double-beta decay end-point
\item[-] Cosmogenics isotopes from Te cosmic activation
\end{enumerate}

The energy spectra of the aforementioned backgrounds, as well as the expected energy spectrum of the neutrinoless double-beta decay signal for an effective Majorana mass of 100~meV after 5 years of data taking and a Te-loading of 0.5\%, are shown in Figure~\ref{fig:phase2_dbd} (right).\\
After 5~years of data taking and with these expected background and signal rates, SNO+ aims at reaching a sensitivity on the half-life of the neutrinoless double-beta decay in $^{130}$Te of 1.9 $\times$ 10$^{26}$ years, which corresponds to a limit on the effective Majorana mass of 50.6~meV.

\section{Conclusions}

SNO+ will imminently enter a fill campaign with scintillator that will last for several months. 
The data taken in the water phase is being analyzed and preliminary results on nucleon decay searches and solar neutrinos observations are expected within a short timescale. 
With the nominal loading of SNO+ in its tellurium phase is 0.5\% by weight, studies performed on higher concentrations of tellurium show that higher concentrations such as 3\% by weight can be reached without significant losses of light transmission in the detector.
The versatility of SNO+ and its loading technique could thus allow it to increase the concentration of tellurium in the future in order to reach limits on the neutrinoless double-beta half-life or more than 10$^{27}$ years, making it possible to start probing the inverted hierarchy domain of the mass phase space.


\Acknowledgements
We acknowledge the generous support of the Vale and SNOLAB staff.\\
The author is funded by the U.S. Department of Energy and the Nuclear Science and  Security  Consortium.\\
This  material  is  based  upon  work  supported  by  the  Department  of  Energy  National  Nuclear Security  Administration  through  the  Nuclear  Science and  Security  Consortium  under  Award Number DE-NA0003180.
This report was prepared as an account of work sponsored by an agency of the United States Government. Neither the United States Government nor any agency thereof, nor any of their employees,  makes any  warranty,  express  or  limited,  or  assumes  any  legal  liability  or responsibility  for  the  accuracy,  completeness,  or  usefulness  of  any  information,  apparatus, product,  or  process  disclosed,  or  represents  that  its  use  would  not  infringe  privately  owned rights.
Reference herein to any specific commercial product, process, or service by trade name, trademark,  manufacturer,  or  otherwise  does  not  necessarily  constitute  or  imply  its endorsement,  recommendation,  or  favoring  by  the United States Government or any agency thereof.
The views and opinions of authors expressed herein do not necessarily state or reflect those of the United States Government or any agency thereof.



\end{document}